\documentclass[10pt]{amsart}
\usepackage{graphicx,studia}

\coordinates{XLVI}{1}{2001} 

\title{Coding objects related to Catalan numbers}

\author{Antal Bege \and Zolt\'an K\'asa}
\address{Babe\c s-Bolyai University, Faculty of Mathematics 
and Computer Science,\newline
RO 3400 Cluj-Napoca, Str. Kog\u alniceanu 1, Romania}
\email{bege@math.ubbcluj.ro, kasa@cs.ubbcluj.ro}
\date{November 1, 2001} 

\subjclass{11B75, 68P30, 68R05} 
\subjclassCR{G.2.1 [\textbf{Mathematics and Computing}]: 
Discrete Mathematics -- \textit{Combinatorics, Counting problems};
} 

\begin{document}

\begin{abstract}
A coding method using binary sequences is presented for different computation 
problems related to Catalan numbers. This method proves in a very easy 
way the equivalence of these problems.
\end{abstract}

\maketitle

\section{Introduction}
The Catalan numbers, named after the french mathematician E. C. Catalan, 
defined as \[ C_n=\frac{1}{n+1}{{2n}\choose n},\]
are as known as the Fibonacci numbers.\footnote[0]{Research supported 
by Sapientia Foundation: \texttt{http://www.sapientia.ro}.}
These numbers arise in a  lot of 
combinatorial problems as the number of some objects. The Catalan number  
$C_n$ describe, among other things,
\begin{itemize}
\item the number of binary trees with $n$ nodes, 
\item the number of ways in which parantheses can be placed in a sequence of
   $n+1$ numbers to be multiplied two at a time,
\item the number of well-formed reverse Polish expressions with $n$ operands and $n+1$ operators,
\item the number of paths in a grid from $(0,0)$ to $(n,n)$, increasing just one coordinate by one
     at each step, without crossing the main diagonal,
\item  the number of $n$-bit sequences that the number of 1s never exceeds the
  number of 0s in each position from left to right,
\item the number of ways you can draw non-crossing segments between $2n$ points 
on a circle in the plane,
\item the number of sequences $(x_1, x_2, \ldots, x_{2n})$, with
 $x_i\in \{-1, 1\}$ for all $i$ between 1 and $2n$ and having the following 
properties for all partial sums:
  $x_1\ge 0, \;x_1+x_2\ge 0, \;\ldots, \; x_1+x_2+\ldots + x_{2n-1}\ge 0, 
\; x_1+x_2+\ldots + x_{2n}= 0$,
\item the number of ways a polygon with $n+2$ sides can be cut into $n$ triangles,
\item the number of frieze pattern with $n+1$ rows,
\item the number of mountain ranges  you can draw using $n$ upstrokes and $n$ downstrokes,
\item the number of ways $n$ votes can come in for each of two candidates in an election, with the 
first never behind the second.
\end{itemize}

The Catalan numbers are the solution of the following recurrence equation:
\[
C_{n+1}=C_0C_n+C_1C_{n-1}+\ldots + C_nC_0 \quad \textrm{ for }  n\ge 0,
\textrm{ with } C_0=1.
\]
 Another recurrence equation for the Catalan numbers is:
\[
(n+2)C_{n+1}=(4n+2)C_n\quad\textrm{ for } n\ge 0,\textrm{ with }C_0=1.\]
The generating function of these numbers is
\[
  \sum_{n\ge 0}{C_nz^n= \frac{1-\sqrt{1-4z}}{2z}},
\]
which can be obtained from the first recurrence equation given above using 
generating function techniques (see e. g. \cite{knu} for computing the number 
of $n$-node binary trees). 

Let $C(z)=\displaystyle\sum_{n\ge 0}{C_nz^n}$ be the generating function 
corresponding to the Catalan numbers. By the recurrence equation this function
satisfy the following equation:
\[
  zC^2(z)=C(z)-1, \quad \textrm{ with } C(0)=1.
\]
From this 
\[
  C(z)= \frac{1-\sqrt{1-4z}}{2z}
\]
results. By developping in series we will get the followings:
\[ 
 C(z)=\frac{1}{2z}\left( 1-\sqrt{1-4z} \right)= \frac{1}{2z}\left(
     1-\sum_{n\ge 0}{{{\frac{1}{2}}\choose n }(-4z)^n}   \right)=
\]
\[
=\sum_{n\ge 0}{{{\frac{1}{2}}\choose {n+1}}(-1)^n 2^{2n+1}z^n}=
\sum_{n\ge 0}{\frac{1}{n+1}{{2n}\choose n}},
\]
and from this the formula for Catalan numbers results.

\section{The Encoding}
We shall present here an encoding method of objects whose number is a 
Catalan number. Each object will be codified by a binary sequence in which 
the number of 0s is equal to the number of 1s, and from the
beginning to any position of the sequence, the number of 1s never exceeds the 
number of 0s. Let us call these sequences \emph{Catalan sequences}. 

The mathematical definition of the Catalan sequence is given below.
Let us denote by $n_1(u)$ the number of 1s and by $n_0(u)$ the number 
of 0s in the sequence $u$. 
A sequence $u=u_1u_2\ldots u_{2n}$, with $u_i\in \{0,1\}$ 
for $i=1,2,\ldots, 2n$, is a \emph{Catalan sequence} if
\begin{eqnarray*}
   n_1(u_1u_2\ldots u_i) &\le& n_0(u_1u_2\ldots u_i) 
             \quad\textrm{ for }\quad i=1,2,\ldots ,2n \\
n_1(u)&=&n_0(u) 
\end{eqnarray*}

Our coding method is different from the one given in \cite{wei} for binary 
trees.

There are a lot of papers which deal with the Catalan numbers, in references 
we give only a few of them.

\subsection{Encoding of binary trees}
The encoding of a binary tree is the following: when a vertex has only one
descendant, we put the sequence 01 for a single left edge, 10 for a single 
right edge, and 00 for the left edge resp. 11 for the right edge when there 
are two descendants. We complete the resulting sequence  with 0 at the 
beginning and 1 at the end. The encoding is made using a preorder 
traversal of the binary tree. In the case of the binary trees with 3 nodes 
we shall have the encoding in Fig. 1. 
 
\begin{figure}
\includegraphics[scale=0.7]{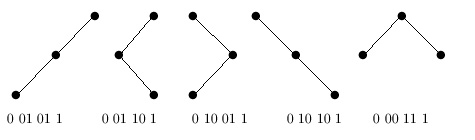}
\caption{Encoding of binary trees}
\end{figure}

\begin{figure}
\includegraphics[scale=0.7]{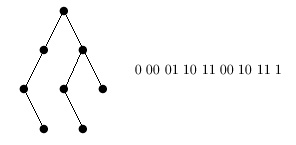}
\caption{A more complex example}
\end{figure}

A more complex example, when the preorder traversal can be easily seen,
is given in Fig. 2.

\bigskip
The encoding algorithm for a binary tree $B$ is given as follow in a 
pseudocode-form. Let us denote by $\emptyset$ the empty binary tree (with no 
vertices). 
The \textbf{put} statement puts its argument in the 
resulting output sequence. 

\newpage
\noindent\emph{Algorithm for encoding a binary tree}

\noindent\textbf{put} 0

\noindent\textbf{procedure} encoding ($B$):

\emph{Let $B_L$ be the left and $B_R$ the right subtree of $B$} 

\textbf{if} $B_L \ne \emptyset$ and $B_R = \emptyset$ \textbf{then} 

\hspace*{4 cm} \textbf{put} 01 

\hspace*{4 cm} \textbf{call} encoding ($B_L$) 

\textbf{if} $B_L=\emptyset$ and $B_R \ne \emptyset$ \textbf{then} 

\hspace*{4 cm} \textbf{put} 10 

\hspace*{4 cm} \textbf{call} encoding ($B_R$) 

\textbf{if} $B_L\ne \emptyset$ and $B_R\ne \emptyset$ \textbf{then} 

\hspace*{4 cm} \textbf{put} 00 

\hspace*{4 cm} \textbf{call} encoding ($B_L$) 
 
\hspace*{4 cm}  \textbf{put} 11

\hspace*{4 cm} \textbf{call} encoding ($B_R$) 

\noindent\textbf{end procedure}

\noindent\textbf{put} 1

\bigskip
For an empty binary tree the procedure has no effect. The proof that the
resulting sequence is a Catalan sequence is immediately by the above 
algorithm.

\subsection{Encoding of paths in grid}
We shall put 0 for a horizontal unit of the path and 1 for a vertical one.
The resulting sequence is a Catalan one because the path never cross the 
main diagonal of the grid.

In the case of a grid $3\times 3$ the following paths and codes results 
(see Fig. 3). 

\begin{figure}
\includegraphics[scale=0.7]{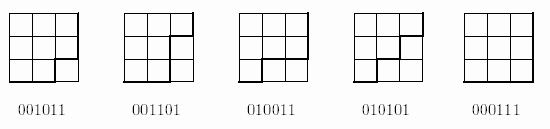}
\caption{Encoding of paths in a grid}
\end{figure}

\subsection{Encoding of expressions with multiplications}

To encode expressions we first attach to each expression for 
multiplication a binary tree by a very simple method. If we multiple 
$a$ by $b$, this yields a binary tree with a root and two 
descendant nodes $a$ and $b$. A multiplication of two expressions
yields a binary tree with two subtrees which are the binary trees 
corresponding to the two expressions. In the resulting binary tree 
each internal nodes has exactly two descendants. Such trees are called
\emph{extended binary trees}. 
To encode an extended binary tree we shall omit all leaves (with of course 
the corresponding edges) in the tree corresponding to the multiplication 
expression and use the encoding method presented before for the resulting 
binary tree. For $n=4$ we shall have the expressions and the corresponding 
extended binary trees in Fig. 4. If we omit all leaves with the adjacent edges 
in these extended trees the binary trees and the corresponding encoding 
result. 

\begin{figure}
\includegraphics[scale=0.7]{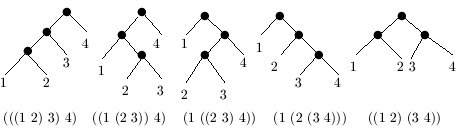}
\caption{Encoding of multiplications}
\end{figure}

\subsection{Encoding of sequences}
We encode sequences $(x_1, x_2, \ldots, x_{2n})$, with
 $x_i\in \{-1, 1\}$ for all $i$ between 1 and $2n$ and having the following 
properties for all partial sums:
  $x_1\ge 0, \;x_1+x_2\ge 0, \;\ldots, \; x_1+x_2+\ldots + x_{2n-1}\ge 0, 
\; x_1+x_2+\ldots + x_{2n}= 0$.
We shall code $-1$ in the sequence by 1 and 1 by 0. It is easy to see 
that in any positions the number of 1s never exceeds the number of 0s, 
and they are equals in the sequence (because the sum of all $2n$ elements 
is equal to 0), so the resulting sequence is a Catalan one. For example:

\medskip
$\begin{array}{rrrrrrrr}
1, & 1  & 1, & -1,& -1,& -1 & \textrm{ coded by: } & 000111\\
1, & 1  & -1,& 1, & -1,& -1 & \textrm{ coded by: } & 001011\\
1, & 1  & -1,& -1,& 1, & -1 & \textrm{ coded by: } & 001101\\
1, & -1 & 1, & 1, & -1,& -1 & \textrm{ coded by: } & 010011\\
1, & -1 & 1, & -1,& 1, & -1 & \textrm{ coded by: } & 010101
\end{array}
$

\subsection{Encoding of segments}
If we have $2n$ points on a circle in the plane and $n$ non-crossing 
segments between them, the encoding is the following: Let us mark the points 
clockwise on the circle with numbers from 1 to $2n$. For a segment between 
$i$ and $j$ ($i<j$) put 0 in the 
$i^{\textrm{th}}$ position and 1 in the $j^{\textrm{th}}$ position in the 
code sequence. For $n=3$ see Fig. 5. It is easy to see that the resulting
sequence is a Catalan one.

\begin{figure}
\includegraphics[scale=0.7]{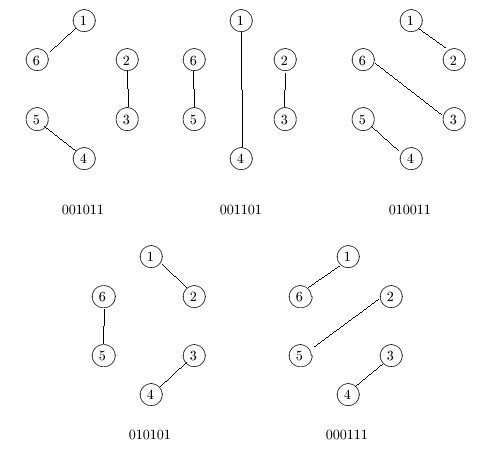}
\caption{Encoding of segments}
\end{figure}

\subsection{Encoding of reverse Polish expressions}
We shall code each operand by 0 and each operator by 1, and add at the 
end of the resulting sequence an 1. For example, if we have the reverse 
Polish expression $aaa\times a \times \times$ --- which corresponds to 
the expression $(a\times ((a \times a)\times a))$ --- the resulting code 
is $00010111$.

\subsection{Encoding of polygons} The polygon is divided into triangles. 
We consider one node in each triangle, and one outside of each side of the
polygon. Join by an edge two nodes if the corresponding triangles (or a 
triangle and the outside of the polygon) have a side in common. We shall
get a tree, on which the encoding will be made. If we mark one side of 
the polygon and the corresponding edge of the tree, and eliminate all edges 
from the tree that have an endpoint as a leave, we
shall get a binary tree (the root will be the node which is adjacent with the 
marked edge). The exemplification will be made for $n=3$ (pentagon). The
marked side is $AB$. (See Fig. 6)

\begin{figure}
\includegraphics[scale=0.7]{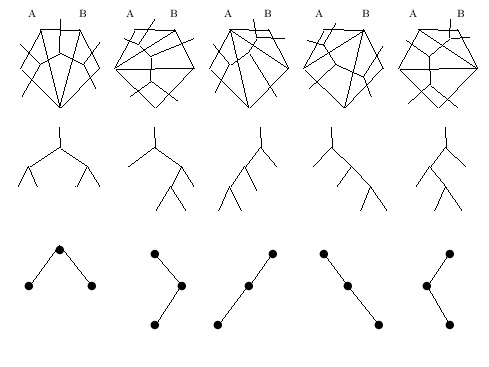}
\caption{Encoding of polygons}
\end{figure}

\section{The Decoding}

If we have a Catalan sequence, from this the corresponding object can be 
easily obtained. Let us consider for exemplification the sequence 
$00010111$.

If we want to obtain the corresponding binary tree, we shall omit the first
0 and the last 1. The subsequence 00 is for a left edge (in a stack we shall 
keep its position), the following subsequence is 10 corresponding to a single 
right edge, the remaining subsequence 11 is a right edge (corresponding to the 
edge kept in the stack). The binary tree obtained is in Fig. 7.a.

For the segments we search the first subsequence 01, trace the corresponding
segment, omit it from the sequence and continue with the remaining sequence 
(keeping the original positions) (Fig. 7.b).

For the multiplication we first draw the corresponding binary tree (Fig. 7.a),
complete it to having two descendants for each node. The resulting extended 
binary tree give us the order of multiplications (Fig. 7.c). 

The path in the grid is obtained immediately: we draw a horizontal unit 
segment for each 0 and a vertical one for each 1 (Fig. 7.d).

From these examples general algorithms to obtain related objects from 
the Catalan sequences can be easily given. We shall describe
only the algorithm to obtain a binary tree from a Catalan sequence.

\begin{figure}
\includegraphics[scale=0.7]{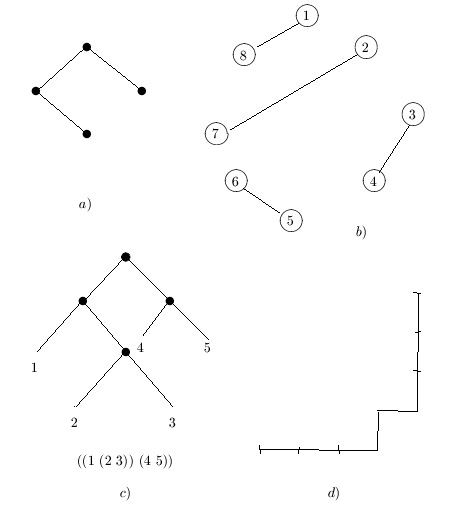}
\caption{Decoding}
\end{figure}

\bigskip
The following recursive algorithm is to decode a Catalan sequence in a 
binary tree. This algorithm is valid only for correct Catalan sequences.
The \textbf{get} statement gets the next two digits from the 
sequence. We shall use the notion of current vertex to denote a vertex 
from which an edge is drawn. After drawing an edge from the current vertex 
the adjacent new vertex will be the current one.

\newpage
\noindent\emph{Algorithm to decode a Catalan sequence into a binary tree}

\medskip
\noindent \textbf{Input} a Catalan sequence

\noindent \textbf{Output} a binary tree

\smallskip
\noindent\textbf{delete} 0 from the beginning and 1 from the end 
of the input sequence, 

\hspace*{1.5 cm} and draw a vertex (the root of the tree) as 
current vertex

\noindent\textbf{procedure} decoding ($c$):

\textbf{get} $ab$ 

\textbf{delete} $ab$ from $c$  

\textbf{if} $ab = 01$ \textbf{then} 

\hspace*{3 cm} draw a left edge from the current vertex 

\hspace*{3 cm} \textbf{call} decoding ($c$) 

\textbf{if} $ab = 10$ \textbf{then} 

\hspace*{3 cm} draw a right edge from the current vertex 

\hspace*{3 cm} \textbf{call} decoding ($c$) 

\textbf{if} $ab = 00$ \textbf{then} 

\hspace*{3 cm} put in the stack the position of the current vertex 

\hspace*{3 cm} draw a left edge from the current vertex
 
\hspace*{3 cm} \textbf{call} decoding ($c$) 

\textbf{if} $ab = 11$ \textbf{then} 

\hspace*{3 cm} get from the stack the position of a vertex, 

\hspace*{3 cm} this will be the current vertex

\hspace*{3 cm} draw a right edge from the current vertex 

\hspace*{3 cm} \textbf{call} decoding ($c$) 

\noindent\textbf{end procedure}

\section{Conclusions}

Our presentation give a uniform method to encode objects whose number is 
a Catalan number. The resulting code is a so-called Catalan sequence
formed of equal number of binary digits 0 and 1, in which the number of 1s 
never exceeds the number of 0s from left to right. This method is important, 
beside the easy handling,   
because coding an object in a Catalan sequence and after decoding it in 
another kind of object, the equivalence of these problems can be easily seen.
To prove that the number of objects in a class is a Catalan number it is 
enough to use the encoding method to obtain a Catalan sequence.

\end{document}